\documentclass[preprint]{revtex4}

\usepackage{amsmath}

\begin{document}

\title{Disentanglement and Decoherence without dissipation at non-zero temperatures}
\author{G. W. Ford}
\affiliation{Department of Physics, University of Michigan, Ann Arbor, MI 48109-1120 USA}
\author{R. F. O'Connell}
\affiliation{Department of Physics and Astronomy, Louisiana State University, Baton
Rouge, LA 70803-4001 USA}
\date{\today}

\begin{abstract}
Decoherence is well understood, in contrast to disentanglement. According to common lore, irreversible coupling to a dissipative environment
is the mechanism for loss of entanglement. Here, we show that, on the
contrary, disentanglement can in fact occur at large enough temperatures $T$
even for vanishingly small dissipation (as we have shown previously for
decoherence). However, whereas the effect of $T$ on decoherence increases
exponentially with time, the effect of $T$ on disentanglement is constant
for all times, reflecting a fundamental difference between the two
phenomena. Also, the possibility of disentanglement at a particular $T$
increases with decreasing initial entanglement.
\end{abstract}

\maketitle

Entanglement, which describes correlations between two or more particles or
subsystems, is an essential characteristic of quantum mechanics and plays a
key role in all applications related to information science \cite
{knight05,amico08,vedral08,horodecki}. But entanglement is poorly understood, so here we attempt to learn more about it by carrying out an exact calculation for the simplest non-trivial system and we will contrast our results with those from an analogous calculation which we already carried out for decoherence. In common with decoherence
(which can occur for just a single particle in a superposition state), entanglement
can be destroyed by interaction with a dissipative heat bath. But, motivated
by the fact that we have previously shown that decoherence can actually
occur at non-zero temperatures $T$ for vanishingly small dissipation 
\cite{ford01, ford02}, our purpose here is to show how disentanglement is
affected by $T$.

In a previous communication \cite{ford}, which was concerned with comparison
of entanglement measures, we considered an entangled system in the absence
of a heat bath and at zero temperature. We now extend our analysis of this
model to incorporate non-zero temperatures and we present an exact
calculation showing that disentanglement can in fact occur in the absence of
dissipation. As we emphasized previously \cite{ford01,ford02}, the situation
is like that for an ideal gas in that collisions (dissipation) are necessary
to bring about an approach to equalibrium but weak enough so that they do
not appear in the equation of state nor in the velocity distribution.

Before proceeding, we should perhaps remark that our method contrasts with the usual master equation approaches where, in general, one starts with an initially uncoupled quantum state, a free particle, say.  Thus, the free particle is essentially at zero temperature with no cognizance of even the zero-point oscillations of the electromagnetic field.  In addition, the initial state of the heat bath is in equilibrium at some temperature $T$ but not coupled to the free particle.  Next, the free particle and heat bath are brought into contact and, as we have shown explicitly \cite{ford01PRD}, the free particle receives an initial impulse with the result that the center of the wave packet drifts to the origin.  But, since for a free particle the origin cannot be a special point, we see that the translational invariance of the problem is broken by the assumption that the initial state corresponds to an uncoupled system.  This problem exists in so-called "exact" master equation formulations, which are exactly only in the sense that they incorporate time-dependent coefficients but they suffer from the same defects as the more conventional master equations; in fact, the same results arise more easily from the use of the initial value Langevin equation which enabled us to obtain solutions of these "exact" master equations in a much more simplified form than one finds in the literature \cite{ford01PRD}.

As in \cite{ford}, we consider two free particles, each of mass $m$, at
positions $x_{1}$ and $x_{2}$, in an initially entangled Gaussian state, but
we extend our previous analysis to allow both particles 1 and 2 to have
velocities $v_{1}$ and $v_{2}$, respectively, which we will eventually take
to be the random velocities associated with a bath at temperature $T$. Thus,
we are dealing with a system with continuous degrees of freedom applicable
to particle position or momenta or to the field modes of light (of interest
in connection with linear optical quantum computing).

The most general initial Gaussian wave function that is symmetric in the two
particles has the form 
\begin{equation}
\psi (x_{1},x_{2};0)=\frac{(a_{11}^{2}-a_{12}^{2})^{1/4}}{\sqrt{2\pi }}\exp
\left\{ -\frac{a_{11}x_{1}^{2}+2a_{12}x_{1}x_{2}+a_{11}x_{2}^{2}}{4}+i\frac{m
}{\hbar }\left( v_{1}x_{1}+v_{2}x_{2}\right) \right\} .  \label{dwod1}
\end{equation}
In order that this state be square-integrable we must of course assume that $
a_{11}$ is positive and that $a_{11}^{2}-a_{12}^{2}>0$. With this wave
function we find the following expressions for the initial correlations
\begin{eqnarray}
\left\langle x_{1}^{2}(0)\right\rangle  &=&\left\langle
x_{2}^{2}(0)\right\rangle =\frac{a_{11}}{a_{11}^{2}-a_{12}^{2}},  \notag \\
\left\langle x_{1}\left( 0\right) x_{2}\left( 0\right) \right\rangle  &=&-
\frac{a_{12}}{a_{11}^{2}-a_{12}^{2}},  \notag \\
\left\langle p_{1}^{2}(0)\right\rangle  &=&m^{2}v_{1}^{2}+\frac{\hbar ^{2}}{4
}a_{11},  \notag \\
\left\langle p_{2}^{2}(0)\right\rangle  &=&m^{2}v_{2}^{2}+\frac{\hbar ^{2}}{4
}a_{11},  \notag \\
\left\langle p_{1}\left( 0\right) p_{2}\left( 0\right) \right\rangle  &=&
\frac{\hbar ^{2}}{4}a_{12},  \notag \\
\frac{\left\langle x_{1}\left( 0\right) p_{1}\left( 0\right) +p_{1}\left(
0\right) x_{1}\left( 0\right) \right\rangle }{2} &=&\frac{\left\langle
x_{2}\left( 0\right) p_{2}\left( 0\right) +p_{2}\left( 0\right) x_{2}\left(
0\right) \right\rangle }{2}=0,  \notag \\
\left\langle x_{2}\left( 0\right) p_{1}\left( 0\right) \right\rangle 
&=&\left\langle x_{1}\left( 0\right) p_{2}\left( 0\right) \right\rangle =0.
\label{2}
\end{eqnarray}

These results are standard quantum mechanics. Next, we consider an ensemble
of particles in thermal equilibrium at a temperature $T$, and so we regard $v_{1}$ and $v_{2}$ as \underline{random} velocities generated by thermal motion.  Also, we consider that the particles are so weakly coupled to a heat bath
that we can neglect dissipation in the equation of motion. In order to ensure that a normalizable thermal state exists for our field - free Hamiltonian, we first put the two particles in an oscillator potential and later take the limit of negligibly small oscillator frequency.  Noting that the initial correlations have no linear terms in the velocities but simply have quadratic terms, we thus obtain the
corresponding expressions by averaging over our thermal distribution of
initial velocities such that
\begin{equation}
v_{1}^{2}\rightarrow \frac{kT}{m},\quad v_{2}^{2}\rightarrow \frac{kT}{m}.
\label{3}
\end{equation}
With this in the expressions (\ref{2}) we have
\begin{equation}
\left\langle p_{1}^{2}(0)\right\rangle =\left\langle
p_{2}^{2}(0)\right\rangle =mkT+\frac{\hbar ^{2}}{4}a_{11}.  \label{4}
\end{equation}
To obtain the time correlations at time $t$, we introduce the time-dependent
(Heisenberg) operators:
\begin{eqnarray}
x_{1}\left( t\right)  &=&x_{1}\left( 0\right) +\frac{p_{1}\left( 0\right) }{m
}t,\quad p_{1}\left( t\right) =p_{1}\left( 0\right) ,  \notag \\
x_{2}\left( t\right)  &=&x_{2}\left( 0\right) +\frac{p_{2}\left( 0\right) }{m
}t,\quad p_{2}\left( t\right) =p_{2}\left( 0\right) .  \label{5}
\end{eqnarray}
With this it is a simple matter to construct the correlations
\begin{eqnarray}
\left\langle x_{1}^{2}\left( t\right) \right\rangle  &=&\frac{a_{11}}{
a_{11}^{2}-a_{12}^{2}}+\left( \frac{kT}{m}+\frac{\hbar ^{2}}{4m^{2}}
a_{11}\right) t^{2},  \notag \\
\left\langle x_{1}\left( t\right) x_{2}\left( t\right) \right\rangle  &=&-
\frac{a_{12}}{a_{11}^{2}-a_{12}^{2}}+\frac{\hbar ^{2}}{4m^{2}}a_{12}t^{2}, 
\notag \\
\left\langle p_{1}^{2}(t)\right\rangle  &=&\left\langle
p_{2}^{2}(t)\right\rangle =mkT+\frac{\hbar ^{2}}{4}a_{11},  \notag \\
\left\langle p_{1}\left( t\right) p_{2}\left( t\right) \right\rangle  &=&
\frac{\hbar ^{2}}{4}a_{12},  \notag \\
\frac{\left\langle x_{1}\left( t\right) p_{1}\left( t\right) +p_{1}\left(
t\right) x_{1}\left( t\right) \right\rangle }{2} &=&\frac{\left\langle
x_{2}\left( t\right) p_{2}\left( t\right) +p_{2}\left( t\right) x_{2}\left(
t\right) \right\rangle }{2}=\left( \frac{\hbar ^{2}}{4m}a_{11}+kT\right) t, 
\notag \\
\left\langle x_{2}\left( t\right) p_{1}\left( t\right) \right\rangle 
&=&\left\langle x_{1}\left( t\right) p_{2}\left( t\right) \right\rangle =
\frac{\hbar ^{2}}{4m}a_{12}t.  \label{6}
\end{eqnarray}

Next, we address the question of entanglement. Since we are dealing with a
Gaussian state, we can use the necessary and sufficient \ condition of Duan
et al. \cite{duan00}. A zero-mean Gaussian state is fully characterized by
its second moments which, for the symmetric case, can be represented by the
following variance (correlation) matrix \cite{duan00,simon00}
\begin{equation}
\mathbf{M}=\left( 
\begin{array}{cc}
\mathbf{G} & \mathbf{C} \\ 
\mathbf{C} & \mathbf{G}
\end{array}
\right) ,  \label{7}
\end{equation}
where
\begin{eqnarray}
\mathbf{G} &=&\left( 
\begin{array}{cc}
\frac{\left\langle x_{1}^{2}\left( t\right) \right\rangle }{L^{2}} & \frac{
\left\langle x_{1}\left( t\right) p_{1}\left( t\right) +p_{1}\left( t\right)
x_{1}\left( t\right) \right\rangle }{2\hbar } \\ 
\frac{\left\langle x_{1}\left( t\right) p_{1}\left( t\right) +p_{1}\left(
t\right) x_{1}\left( t\right) \right\rangle }{2\hbar } & \frac{
L^{2}\left\langle p_{1}^{2}(t)\right\rangle }{\hbar ^{2}}
\end{array}
\right) ,  \notag \\
\mathbf{C} &=&\left( 
\begin{array}{cc}
\frac{\left\langle x_{1}\left( t\right) x_{2}\left( t\right) \right\rangle }{
L^{2}} & \frac{\left\langle x_{1}\left( t\right) p_{2}\left( t\right)
\right\rangle }{\hbar } \\ 
\frac{\left\langle x_{1}\left( t\right) p_{2}\left( t\right) \right\rangle }{
\hbar } & \frac{L^{2}\left\langle p_{1}(t)p_{2}(t)\right\rangle }{\hbar ^{2}}
\end{array}
\right) .  \label{8}
\end{eqnarray}
and $L$ is a constant of dimension length introduced to make the matrix
elements dimensionless.

In order to discuss entanglement, Duan et al. perform a sequence of
rotations and squeezes to bring $\mathbf{M}$ to a form in which
\begin{equation}
\mathbf{G}=\left( 
\begin{array}{cc}
g & 0 \\ 
0 & g
\end{array}
\right) ,\quad \mathbf{C}=\left( 
\begin{array}{cc}
c & 0 \\ 
0 & c^{\prime }
\end{array}
\right) .  \label{9}
\end{equation}
Since the determinants are invariant under these transformations, we have
the following simple relations for determining the quantities $g$, $c$ and $
c^{\prime }$ in terms of these invariants.
\begin{equation}
\det \mathbf{G}=g^{2},\quad \det \mathbf{C}=cc^{\prime },\quad \det \mathbf{M
}=\left( g^{2}-c^{2}\right) \left( g^{2}-c^{\prime 2}\right) .  \label{10}
\end{equation}
With the expressions (\ref{6}) for the correlations we find
\begin{eqnarray}
\det \mathbf{G} &=&\frac{\left( a_{11}+\frac{4mkT}{\hbar ^{2}}\right) a_{11}
}{4\left( a_{11}^{2}-a_{12}^{2}\right) },  \notag \\
\det \mathbf{C} &=&-\frac{a_{12}^{2}}{4\left( a_{11}^{2}-a_{12}^{2}\right) },
\notag \\
\det \mathbf{M} &=&\left( \frac{1}{4}+\frac{mkT}{\hbar ^{2}\left(
a_{11}-a_{12}\right) }\right) \left( \frac{1}{4}+\frac{mkT}{\hbar ^{2}\left(
a_{11}+a_{12}\right) }\right) .  \label{11}
\end{eqnarray}
Putting these in (\ref{10}) and solving, we find
\begin{eqnarray}
g &=&\frac{1}{2}\sqrt{\frac{\left( a_{11}+\frac{4mkT}{\hbar ^{2}}\right)
a_{11}}{\left( a_{11}^{2}-a_{12}^{2}\right) }},  \notag \\
c &=&\frac{\left\vert a_{12}\right\vert }{2}\sqrt{\frac{a_{11}+\frac{4mkT}{
\hbar ^{2}}}{\left( a_{11}^{2}-a_{12}^{2}\right) a_{11}}},  \notag \\
c^{\prime } &=&-\frac{a_{11}\left\vert a_{12}\right\vert }{2\sqrt{\left(
a_{11}^{2}-a_{12}^{2}\right) \left( a_{11}+\frac{4mkT}{\hbar ^{2}}\right)
a_{11}}}.  \label{12}
\end{eqnarray}

Duan et al have obtained a necessary and sufficient condition that a
Gaussian state is separable. In terms of these quantities their condition is
equivalent to the inequality
\begin{equation}
\left( g-c\right) \left( g-c^{\prime }\right) \geq \frac{1}{4}.  \label{13}
\end{equation}
With the expressions (\ref{12}) this becomes
\begin{equation}
\frac{a_{11}-\left\vert a_{12}\right\vert +\frac{4mkT}{\hbar ^{2}}}{
a_{11}+\left\vert a_{12}\right\vert }\geq 1,  \label{14}
\end{equation}
so that
\begin{equation}
\left\vert a_{12}\right\vert \leq \frac{2mkT}{\hbar ^{2}}.  \label{15}
\end{equation}
It should be emphasized that this condition for distanglement is independent
of time. This is in stark contrast with the corresponding result for
decoherence where the temperature effect increases exponentially to the
power of $t^{2}$ \cite{ford01}. Moreover, if $\vert a_{12}\vert ~>~ \frac{2mkT}{\hbar^{2}}$, the system remains entangled for all time.

Our conclusion is that whereas decoherence and disentanglement always occur for the same system in the presence of dissipation, this is not so for neglible dissipation at temperatures such that $\vert a_{12}\vert ~>~ 2mkT/\hbar^{2}$, in which case decoherence still occurs but disentanglement does not.  It is clear that they are very different phenomena but, whereas decoherence is well understood, the opposite is true for disentanglement.

\section*{ACKNOWLEDGMENT}

This work was partially supported by the National Science Foundation under Grant No. ECCS-0757204.

\end{document}